\documentclass[sigconf]{acmart}

\copyrightyear{2026}
\acmYear{2026}
\setcopyright{cc}
\setcctype{by}
\acmConference[MSR '26]{23rd International Conference on Mining Software Repositories}{April 13--14, 2026}{Rio de Janeiro, Brazil}
\acmBooktitle{23rd International Conference on Mining Software Repositories (MSR '26), April 13--14, 2026, Rio de Janeiro, Brazil}
\acmPrice{}
\acmDOI{10.1145/3793302.3793610}
\acmISBN{979-8-4007-2474-9/2026/04}

\usepackage{pifont}
\usepackage{listings}
\usepackage{xcolor}
\usepackage{hyperref}
\usepackage{xcolor}

\definecolor{softblue}{RGB}{173, 216, 230}

\definecolor{codebg}{HTML}{EEEEEE} % A light gray similar to blue!5!gray!5
\newcommand{\code}[1]{%
  \colorbox{codebg}{\texttt{#1}}%
}

\definecolor{headergreen}{RGB}{112, 173, 71}  
\definecolor{bodygreen}{RGB}{235, 245, 235}   

\newsavebox{\answerbody}
\newenvironment{AnswerBox}[1]{%
  \par\medskip\noindent
  \colorbox{headergreen}{%
    \begin{minipage}{\dimexpr\linewidth-2\fboxsep\relax}
      \textcolor{white}{\textbf{\large #1}}
    \end{minipage}%
  }%
  \par\nointerlineskip % Remove gap
  \begin{lrbox}{\answerbody}%
    \begin{minipage}{\dimexpr\linewidth-2\fboxsep\relax}
      \medskip 
}{%
      \medskip 
    \end{minipage}
  \end{lrbox}%
  \noindent
  \colorbox{bodygreen}{\usebox{\answerbody}}
  \par\medskip
}

\begin{document}

\title{Safer Builders, Risky Maintainers: A Comparative Study of Breaking Changes in Human vs Agentic PRs}

\author{K M Ferdous}
\email{kferdous@students.kennesaw.edu}
\affiliation{%
 \institution{Kennesaw State University}
  \city{Marietta}
  \state{Georgia}
  \country{USA}
}

\author{Dipayan Banik}
\email{dipayan5175@gmail.com}
\affiliation{%
  \institution{Quanta Technology}
  \city{Raleigh}
  \state{North Carolina}
  \country{USA}
}

\author{Kowshik Chowdhury}
\email{kchowdh1@students.kennesaw.edu}
\affiliation{%
  \institution{Kennesaw State University}
  \city{Marietta}
  \state{Georgia}
  \country{USA}
}

\author{Shazibul Islam Shamim}
\email{mshamim@kennesaw.edu}
\affiliation{%
  \institution{Kennesaw State University}
  \city{Marietta}
  \state{Georgia}
  \country{USA}
}

\begin{abstract}

AI coding agents are increasingly integrated into modern software engineering workflows, actively collaborating with human developers to create pull requests (PRs) in open-source repositories. Although coding agents improve developer productivity, they often generate code with more bugs and security issues than human-authored code. While human-authored PRs often break backward compatibility, leading to breaking changes, the potential for agentic PRs to introduce breaking changes remains underexplored. The goal of this paper is to help developers and researchers evaluate the reliability of AI-generated PRs by examining the frequency and task contexts in which AI agents introduce breaking changes.

We conduct a comparative analysis of 7,191 agent-generated PRs with 1402 human-authored PRs from Python repositories in the AIDev dataset. We develop a tool that analyzes code changes in commits corresponding to the agentic PRs and leverages an abstract syntax tree (AST) based analysis to detect potential breaking changes. Our findings show that AI agents introduce fewer breaking changes overall than humans (3.45\% vs. 7.40\%) in code generation tasks. However, agents exhibit substantially higher risk during maintenance tasks, with refactoring and chore changes introducing breaking changes at rates of 6.72\% and 9.35\%, respectively. We also identify a “Confidence Trap” where highly confident agentic PRs still introduce breaking changes, indicating the need for stricter review during maintenance oriented changes regardless of reported confidence score.

\end{abstract}


% \ccsdesc[300]{Software and its engineering~Collaboration in software development; Software maintenance}


\maketitle

\section{Introduction}
\label{introduction}

The AI coding agents, such as Devin, Claude Code, and GitHub Copilot, have brought about a paradigm shift in modern software development workflows \cite{treude_2025} \cite{hassan2025_agentic_SE}. These AI agents enhance developer productivity by performing coding tasks, generating test cases, and handling complex end-to-end development tasks, such as issue resolution and pull request (PR) creation  \cite{copilot-productivity} \cite{vaithilingam2022expectation} \cite{jimenez2023swebench} \cite{chen_2021}. These agents have become active participants alongside human developers in creating pull requests in open-source repositories \cite{hassan2025_agentic_SE} \cite{hassan2024_ainative_se}. For instance, OpenAI Codex has created more than 400,000 PRs in open-source GitHub repositories within the first two months of its release \cite{li_2025_riseaiteammates}. Agentic PRs are increasingly common in real repositories, and datasets such as AIDev capture thousands of these contributions across diverse projects \cite{li_2025_riseaiteammates}.

Despite improvements in developer productivity and active involvement in modern software engineering processes, AI-generated code remains susceptible to significant quality issues. The AI generated code often contains more bugs and security vulnerabilities than human-authored code \cite{bui2025assessment} \cite{khoury2023securecode} \cite{pearce2022asleep} \cite{sandoval2023userstudy}. According to a recent study by CodeRabbit of 470 open-source pull requests on GitHub, AI-generated code contains 1.7 times more issues than human-authored code \cite{CodeRabbit2025_AIvsHuman}. Researchers also reported that developers using AI assistance experienced higher rates of broken tests and integration failures during refactoring tasks \cite{Perry_2023}.

While prior work has primarily focused on AI agent performance and bug analysis, comparatively little attention has been given to their impact on backward compatibility. AI agents may unintentionally introduce breaking changes when generating patches and pull requests \cite{Zhang_2020} \cite{Xavier_2017}, yet this risk remains underexplored compared to human-authored changes. In this research, we investigate whether AI agents can introduce breaking changes.

To address this gap, we evaluate the reliability of agent-generated pull requests by analyzing the frequency of breaking changes and the development contexts in which they occur. We also include our replication package to replicate our findings \cite{msr2026:BC}. Our study addresses the following research questions:

\begin{itemize}

\item  \textbf{RQ1:} \textit{How often do AI agents introduce breaking changes compared to human developers?}

\item \textbf{RQ2:} \textit{How does the breaking change rate of AI-generated pull requests differ between generative tasks and maintenance tasks?}

\item \textbf{RQ3:} \textit{To what extent are AI agent confidence scores associated with the occurrence of breaking changes?}

\end{itemize}

\section{Background}
\label{background}

\begin{figure}[h]
    \centering
    \includegraphics[width=0.9\linewidth]{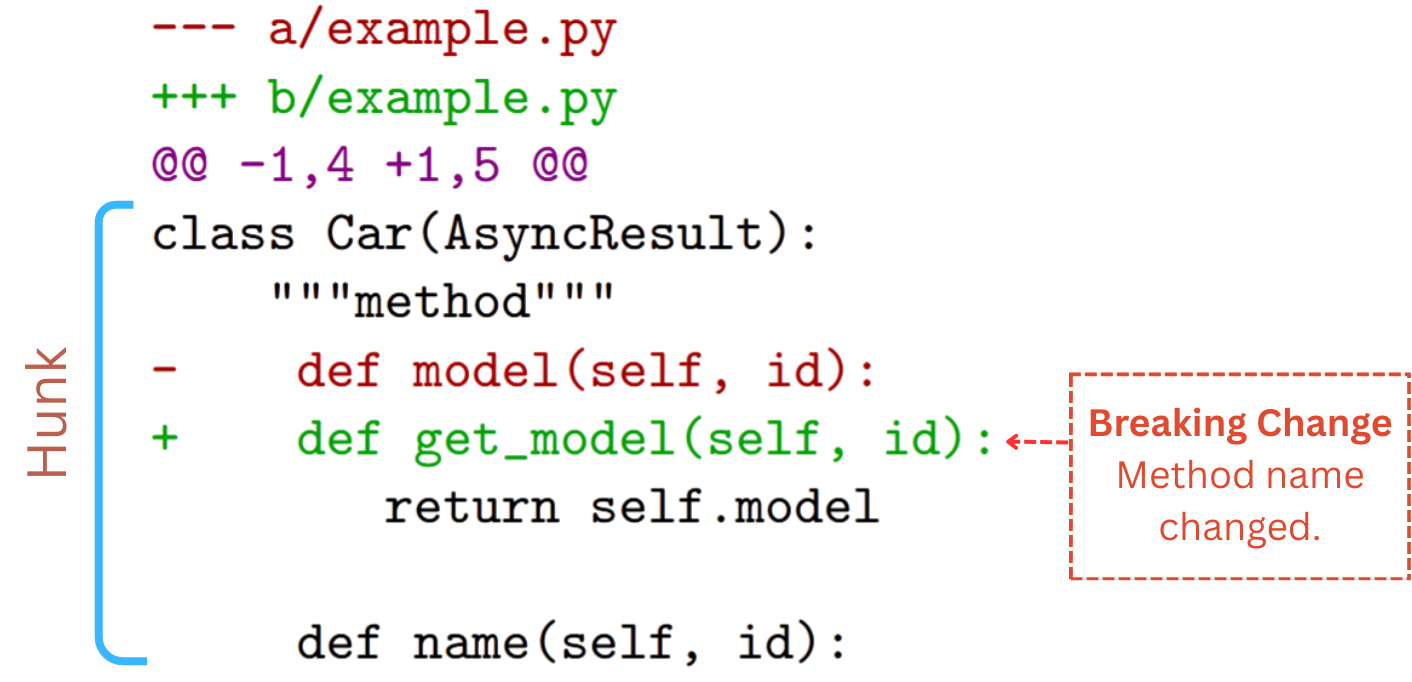}
    \caption{Git hunk and Breaking change}
    \label{fig:git-hunk}
\end{figure}

A breaking change is a code modification that violates backward compatibility and disrupts existing usage \cite{Ochoa_2022}. Such changes often result from structural alterations to a program’s public interface, such as renaming functions or classes, modifying parameters or return types, changing class hierarchies, or removing public members \cite{Brito_2018} \cite{Du_aexpy} \cite{Xavier_2017}. These changes can break dependent code, as illustrated in Figure \ref{fig:git-hunk}.

% A breaking change refers to a modification in software code that is incompatible with its existing usage, potentially violating backward compatibility \cite{Ochoa_2022}. Such changes typically arise from structural modifications that are directly observable in a program’s public interface or API signature \cite{Brito_2018} \cite{Xavier_2017}. Common examples include renaming functions or classes, changing parameter lists, altering class hierarchies, modifying return types, or removing public members \cite{Du_aexpy}. These changes can break dependent code and cause compatibility issues. Figure \ref{fig:git-hunk} illustrates an example of a breaking change.

In Git-based version control systems, code changes are represented as diffs that capture line-level differences between file revisions \cite{gitdiff}. A diff is composed of one or more hunks, each representing a contiguous block of added, removed, or modified lines \cite{GNU_Hunks}. Figure \ref{fig:git-hunk} illustrates a hunk within a Git diff.

% In Git-based version control systems, code changes are represented using diffs, which capture line-level differences between two revisions of a file \cite{gitdiff}. A diff consists of one or more hunks, where a hunk represents a Git diff is a contiguous block of lines in a file that contains changes (additions, deletions, or modifications) \cite{GNU_Hunks}. In figure \ref{fig:git-hunk}  shows a hunk in a git diff structure.

\section{Related works}
\label{related-works}

Breaking changes are a well established concern in software evolution. Xavier et al. \cite{Xavier_2017} studied the frequency of breaking changes, while Zhang et al. \cite{Zhang_2020} found that over 40\% of API changes in Python packages are breaking. Du et al. \cite{Du_aexpy} proposed AexPy, a Python specific tool for detecting breaking changes. However, prior work focuses on human-authored code, leaving AI-generated contributions largely unexplored.

Prior studies on AI programming tools report improvements in productivity \cite{copilot-productivity} and demonstrate the ability to generate complete pull requests \cite{Horikawa2025Agentic} \cite{li_2025_riseaiteammates}, but also reveal limitations in correctness and contextual understanding \cite{vaithilingam2022expectation}. Li et al. \cite{li_2025_riseaiteammates}, examining the transition to “Software Engineering 3.0” through the AIDev dataset, highlight a gap between benchmark performance and real-world effectiveness. They find that AI agents increase contribution volume but achieve lower acceptance rates than human developers. Pearce et al. \cite{pearce2022asleep} found that up to 40\% of AI-generated code may contain security vulnerabilities, while Chen et al. \cite{chen_2021} reported that only 28–48\% of AI-generated solutions pass all tests, indicating frequent failures. Additionally, Horikawa et al. \cite{Horikawa2025Agentic} observed that over 53\% of agentic refactoring occurs implicitly within commits focused on other tasks. 

Although prior research benchmarks AI performance, there is limited research on detecting breaking changes in AI-generated contributions. Our work addresses this gap by comparing task-wise breaking change rates between AI agent-generated and human-authored pull requests and evaluating whether agent confidence scores predict breaking changes.

\section{Methodology}
\label{methodology}

To address our research questions, we use the AIDev dataset \cite{AIDev_HuggingFace}, which provides a comprehensive collection of both AI-agent and human-authored contributions. Our analysis focuses on the patches (Git diffs) included in the dataset. We restrict our study to pull requests from Python-based repositories, as Python is heavily represented in LLM training corpora \cite{kocetkov_2022, chen_2021}, ensuring that our results accurately reflect the capabilities of AI coding agents.

The \code{repository} table lists 2,807 GitHub repositories with at least 100 stars, including 530 Python projects. From these repositories, the \code{pull\_request} table contains 7,191 AI-generated pull requests and 1,402 human-authored pull requests.

\begin{figure}[h]
    \centering
    \includegraphics[width=0.8\linewidth]{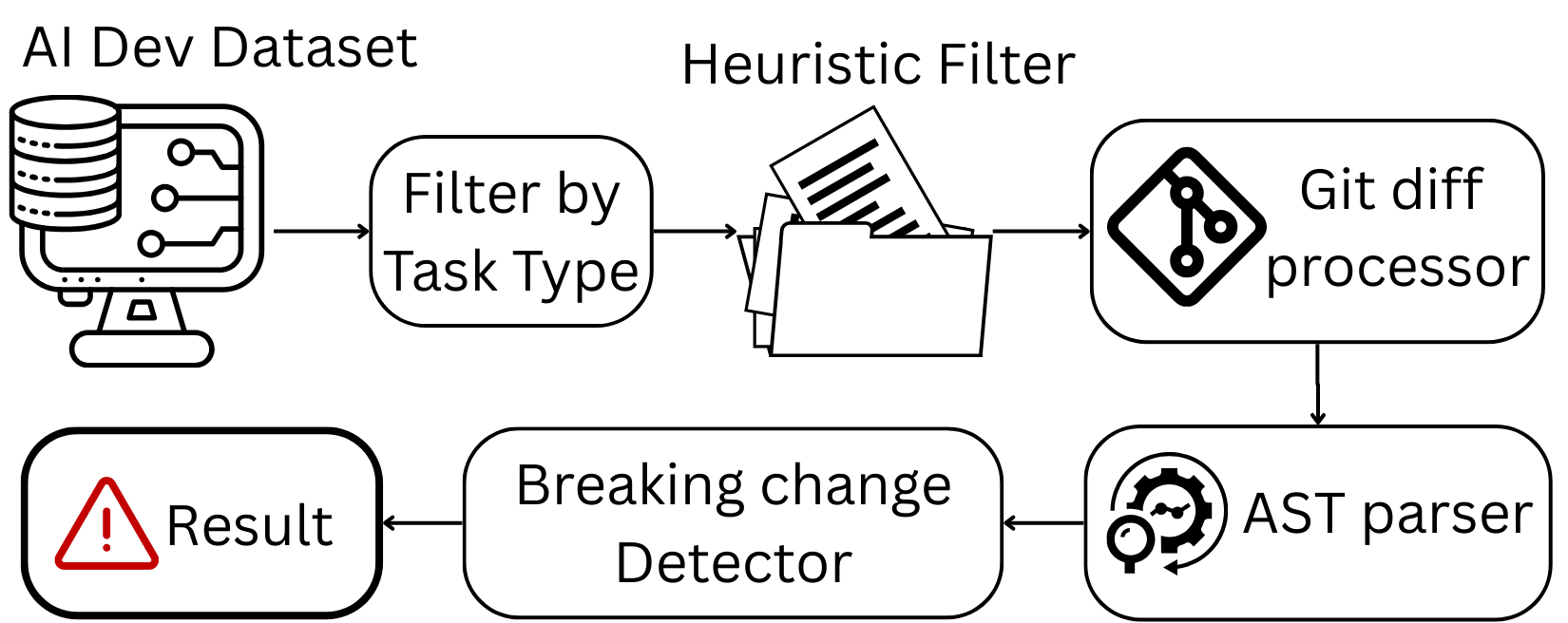}
    \caption{Workflow of potential Breaking change detection}
    \label{fig:workflow}
\end{figure}

 Since state of the art AexPy \cite{Du_aexpy} and other tools like pidiff \cite{Rohanpm_PIDiff}, PyCompat \cite{Zhang_2020} analyze differences between full source code versions rather than directly parsing Git diffs, making these tools unsuitable for our study. Therefore, we develop a tool to detect potential breaking changes from Git diffs. Figure \ref{fig:workflow} illustrates the workflow of our breaking change detection process.

In the first stage, we filter pull requests by selecting five task categories: feat, fix, perf, refactor, and chore, which directly impact program structure according to the commit convention \cite{Angular_CommitGuidelines, ConventionalCommits_Org}. This step results in 4,798 AI-generated pull requests. Using tables \code{pr\_commits} and \code{pr\_commit\_details}, we extract commit-level metadata, yielding 75,467 file-level patches.

Applying the same criteria to human-authored pull requests, we identify 1,026 relevant PRs. Although the dataset includes tables \code{human\_pull\_request} and \code{human\_pr\_task\_type}, it does not provide commit-level data and file patches. To address this limitation, we used a GitHub API–based mining script to retrieve the missing commit data, collecting 5,788 commits and 93,044 file-level patches for analysis.

We then applied a heuristic path-based filter to keep only core Python source files and exclude non-essential directories (e.g. tests, examples, etc). This process results in 23,333 agentic and 36,991 human-authored patches (60,324 total) and removing approximately 66\% of the original patches.

% Then, we apply a heuristic path-based filter to retain only core Python source files (e.g., .py) and exclude non-essential directories such as tests, examples, documentation, and build files. This process results in 23,333 agentic patches and 36,991 human-authored patches, totaling 60,324 patches and removing approximately 66\% of the total patches.

Since AST parser cannot process a patch (git diff) directly, in this step, we reconstruct the pre-commit and post-commit versions of code for each diff hunk using the git diff processor. These reconstructed codes contain valid syntax suitable for AST parsing.

Finally, we parse the pre-commit and post-commit code of all 60,324 patches using Python’s AST parser, categorizing changes by scope like class-level, function-level, etc. The breaking change detector then applies 17 patterns from Du et al. \cite{Du_aexpy} to identify potential breaking changes, organized into three categories: Removals, Modifications, and Additions.

% Then the breaking change detector identifies potential breaking changes by applying the 17 breaking change patterns proposed by Du et al. \cite{Du_aexpy} and produces the results that highlight the patches that contained potential breaking changes. These patterns organized into three categories: (i) \textbf{Removals}, involves the deletion of classes, functions, or class attributes; (ii) \textbf{Modifications}, includes renaming classes, functions, or parameters, or reordering function arguments; (iii) \textbf{Additions}, such as inserting a parameter in the middle of an existing argument list. 

We measure the Potential Breaking Change Rate, defined as the percentage of commit files (patches) that contain at least one detected breaking change.

\begin{equation}
\small
\begin{split}
    &\text{Potential Breaking Change (PBC) Rate} = \\
    &\quad \frac{\text{Number of patches contain Potential Breaking Change}}{\text{Total Number of patches}}
\end{split}
\end{equation}

We ran our potential breaking change detector on 60,324 patches and detected 3,538 potential breaking changes. To validate the reliability of our tool, we randomly selected 94 patches from 3,538 detected potential breaking changes using a 95\% confidence level and a 10\% margin of error. Two authors independently reviewed the sample, confirming 90/94 (95.7\%) and 88/94 (93.6\%) true positive matches, respectively. This yielded substantial inter-rater agreement (Cohen’s Kappa = 0.79), and any disagreements were resolved through discussion. 

\section{Analysis and Findings}
\label{results}

\subsection{Frequency of Potential Breaking Changes (AI vs. Humans)}

\begin{figure}[h]
    \centering
    \includegraphics[width=0.57\linewidth, height=3.3cm]{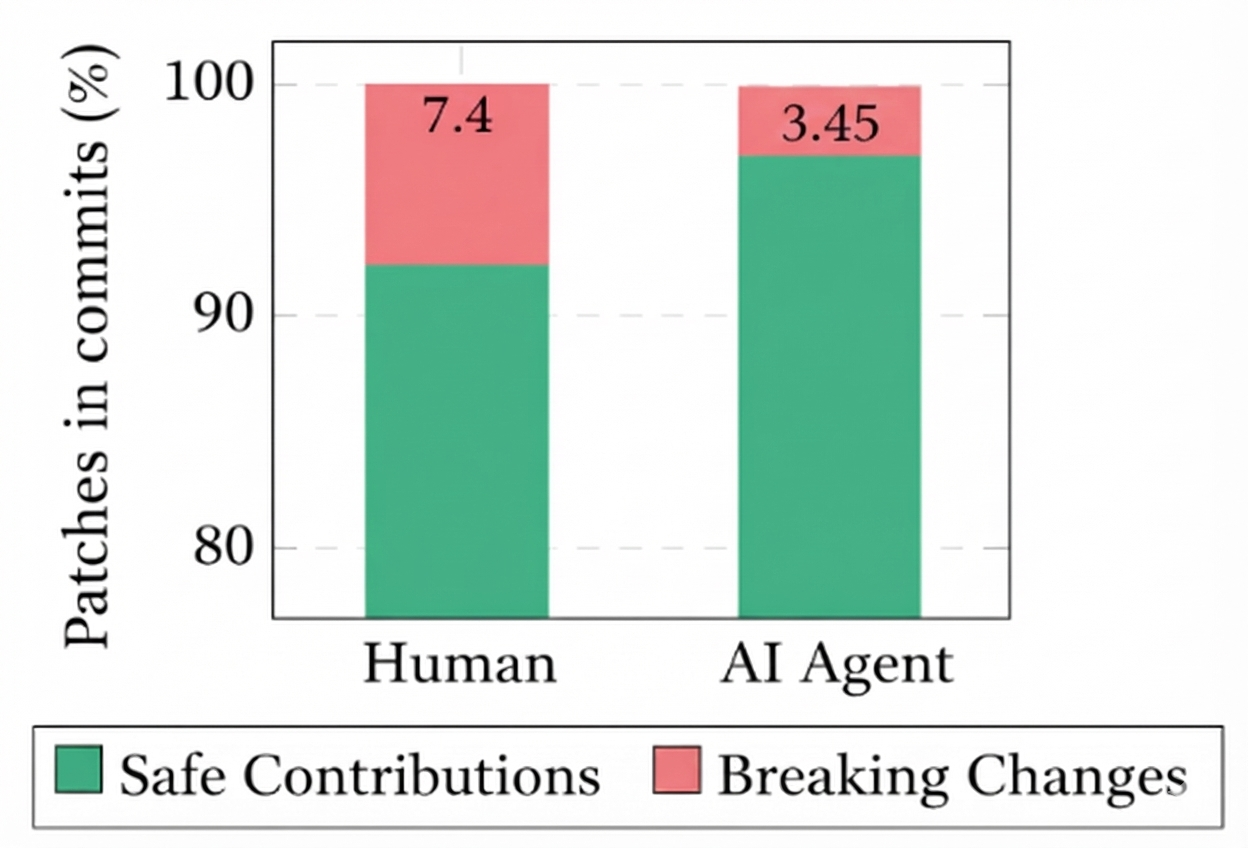}
    \caption{Breaking Change Rates (\%) in AI Agents vs Humans}
    \label{fig:agent-vs-human}
\end{figure}

Our analysis shows that AI agents introduced potential breaking changes in 805 of 23,333 patches (3.45\%), affecting 11.3\% of agent-generated pull requests. In comparison, human developers introduced potential breaking changes in 2,733 of 36,991 patches (7.40\%), impacting 21.18\% of human-authored pull requests. Figure \ref{fig:agent-vs-human} summarizes these results, indicating that AI-generated commits exhibit a lower rate of potential breaking changes than human-authored commits.

\begin{figure}[h]
    \centering
    \includegraphics[width=0.8\linewidth, height=5cm]{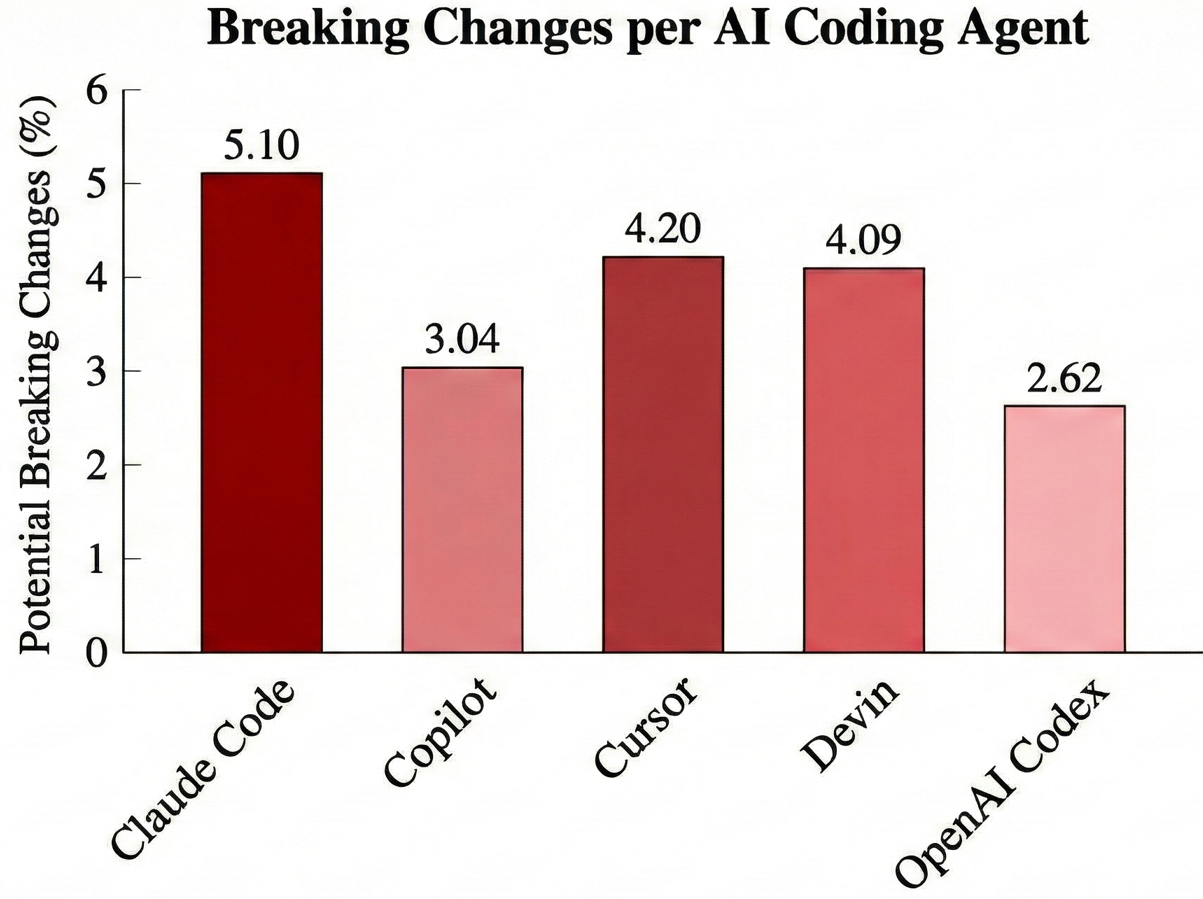}
    \caption{Agent wise potential breaking change rate (\%)}
    \label{fig:result_agent_wise_bc}
\end{figure}

Figure \ref{fig:result_agent_wise_bc} demonstrates the distribution of potential breaking changes across AI agents. Claude Code exhibits 74 breaking changes across 1,450 patches (ratio 5.10), while Copilot, Cursor, Devin, and OpenAI Codex have ratios of 3.04, 4.20, 4.09, and 2.62, respectively. These results indicate that although all agents introduce some potential breaking changes, their frequency remains lower than that observed in human-authored patches.

\begin{AnswerBox}{Answer of RQ1}
Our analysis shows that AI agents introduce potential breaking changes at a lower rate (3.45\%) compare to human developers (7.40\%). While all agents produce some potential breaking changes, their frequency remains lower than that of human-authored commits.
\end{AnswerBox}

\subsection{Task-Specific Breaking Change (Generative vs. Maintenance)}
\label{result-rq2}

\begin{figure}[h]
    \centering
    \includegraphics[width=0.6\linewidth, height=3.7cm]{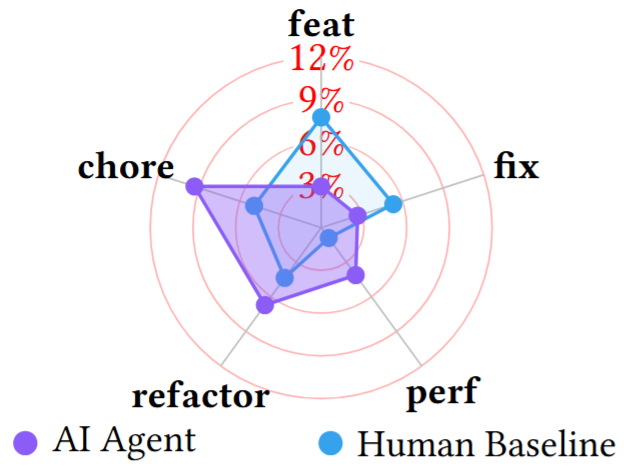}
    \caption{Task-Specific breaking change rate comparison.}
    \label{fig:task-wise-bc}
\end{figure}

To address RQ2, we categorize tasks into generative activities (feat, fix, perf) and maintenance activities (refactor, chore). Figure \ref{fig:task-wise-bc} shows that AI agents exhibit relatively low breaking change rates in generative tasks: 2.89\% for feature additions, 2.69\% for bug fixes, and 4.12\% for performance improvements. In contrast, maintenance tasks carry higher risks, with refactoring at 6.72\% and chore-related tasks at 9.35\%.

For human-authored patches, we notice the opposite trend. Generative tasks have higher breaking change rates, with feat at 7.74\%, fix at 5.32\%, and perf at 0.90\%, whereas maintenance tasks show relatively lower rates, with refactor at 4.36\% and chore at 4.95\%.

Overall, AI agents demonstrate more reliability for code generation but are more risky in maintenance tasks, whereas humans show the opposite trend. This comparison suggests that AI agents need more understanding for maintenance related patches and the changes should undergo careful review.

\begin{AnswerBox}{Answer of RQ2}
AI agents are more prone to breaking changes in maintenance tasks (chore: 9.35\%, refactor: 6.72\%) than in generative tasks (feat: 2.89\%, fix: 2.69\%), indicating that AI agents require deeper understanding for structural modifications.
\end{AnswerBox}

\subsection{The Confidence Trap}
We notice that the confidence scores are strongly right-skewed, with 99.9\% of AI-generated pull requests between 8 and 10. Therefore, we focus on confidence scores 8, 9, and 10 to assess the relationship between confidence and breaking changes.

\begin{figure}[h]
    \centering
    \includegraphics[width=0.6\linewidth, height=3.5cm]{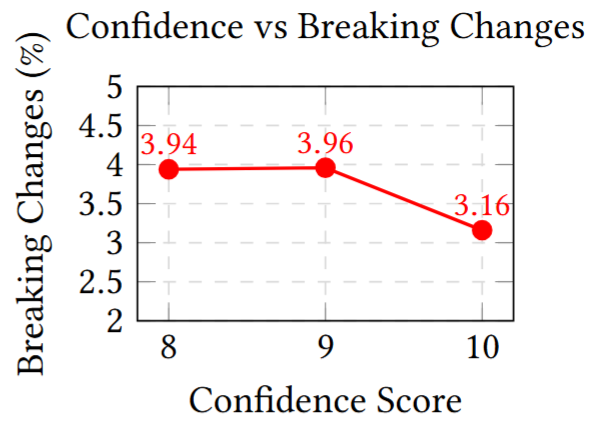}
    \caption{The relationship between AI agent confidence scores and potential breaking changes.}
    \label{fig:confidence-line}
\end{figure}

In Figure \ref{fig:confidence-line}, our findings highlight that breaking change rates are relatively similar for levels 8 and 9, at 3.94\% (33 of 837 patches) and 3.96\% (311 of 7,854 patches), respectively. Interestingly, confidence level 10 also exhibits potential breaking changes, but at a slightly lower rate of 3.16\% (458 breaks out of 14,509 commits).

These results indicate that high confidence does not reliably guarantee safe code generation. Therefore, confidence scores alone are insufficient for prioritizing review or deployment, and should be supplemented with additional verification mechanisms when using agent-generated code in production.

\begin{AnswerBox}{Answer of RQ3}
Breaking changes occur across all high-confidence levels (8–10), ranging from 3.16\% to 3.96\%, indicating that AI agents’ confidence scores alone do not guarantee safe code.
\end{AnswerBox}

\section{Discussion}
\label{discussion}

\subsection{Interpretation of Findings}

\subsubsection{Maintenance Task Risk :}
Our results indicate that AI agents introduce substantially higher rates of potential breaking changes in maintenance oriented tasks, such as refactoring (6.72\%) and chore (9.35\%) related updates (figure \ref{fig:task-wise-bc}). This highlights the need for further research to improve AI agent performance and reliability specifically in maintenance tasks.

\subsubsection{Unreliable Confidence Score :} 
Figure \ref{fig:confidence-line} highlights that AI-generated PRs introduce potential breaking changes even at high confidence levels (8–10), indicating that confidence scores do not reliably reflect breaking change risk. This suggests the need to align confidence with structural risk.

\subsection{Implications for Practitioners}
\subsubsection{Task-Specific Review Policy :}
In figure \ref{fig:task-wise-bc}, we observe that agent-generated maintenance tasks introduce more potential breaking changes than generative tasks. So, we recommend practitioners apply enhanced, task-specific review policies regardless of the agent’s reported confidence.

\subsection{Implications for Researchers}
\subsubsection{Assessing Breaking Changes in Benchmarks :}
Current AI coding benchmarks (e.g., HumanEval, SWE-bench) focus on functional correctness, but agents can introduce breaking changes (Figure \ref{fig:result_agent_wise_bc}). So future research should incorporate breaking-change analysis into these benchmarks.

\section{Threats to Validity}
\label{threat-to-validity}
\subsection{Internal Validity}
Our analysis focuses on five task categories and their PRs that directly affect program structure. This selection may have overlooked “tangled commits”, combining code and documentation but labeled as document type tasks. Additionally, our static analysis may overestimate breaking changes due to challenges in handling nested functions, which are not publicly accessible and difficult to identify from patches \cite{InnerFunctions}.

\subsection{External Validity}
For task-based pull request filtering, we relied exclusively on the classification schema provided by the AIDev dataset. Any misclassifications in the dataset’s tagging logic could affect the transferability of our task-specific findings.

\subsection{Construct Validity}
We measure 'Potential Breaking Changes' based on syntactic-level modifications, even though some changes may affect functions with no downstream users. However, in API evolution, any syntactic-level change is considered a breaking change, irrespective of its usage. \cite{Keshani_2023}. Additionally, our findings are limited to the Python ecosystem, and statically typed languages need further investigation.

\section{Conclusion}
\label{conclusion}

AI coding agents are increasingly used in software development but may introduce breaking changes through unintended structural modifications. In this study, we conduct a comparative analysis of agent-generated and human-authored Python pull requests and find that although agents introduce fewer breaking changes overall, they are significantly more prone to breaking changes during maintenance tasks, especially refactoring and chore-related changes. We also show that agent confidence scores poorly predict breaking change risk. These findings highlight the need for task-aware review processes and new benchmarks that explicitly evaluate breaking change risks in AI-generated code.

\bibliographystyle{ACM-Reference-Format}
\bibliography{references}

\end{document}